\documentstyle[preprint,aps,prd]{revtex}
\begin{document}
%
\preprint{\vbox{\hbox{
 EHU-FT 95/15}\hbox{
 hep-th/9604119}}}
\title{ Bogomolnyi-type bounds in unconventional
superconductors without external magnetic fields}   
\author{Ana Ach\'ucarro \cite{aa}}
\address{Departamento de Fisica T\'eorica,
 Universidad del Pa\'\i s Vasco,
Lejona, Vizcaya, Spain}
\address{Department of Mathematics, Tufts
University, Medford, MA (U.S.A.)}
\address{Departament of Theoretical Physics,
University of Groningen, The Netherlands}
\author{Juan Luis Ma\~nes\cite {jlm}}
\address{Departamento de F\'\i sica de la Materia
Condensada\\ Universidad del Pa\'\i s Vasco, Apartado 644,
48080 Bilbao, Spain } 
\date{December  1995}
\maketitle
\begin{abstract}
Following Bogomolnyi's classical treatment of vortices,
we develop a method for finding rigorous lower bounds to
the Landau-Ginzburg free energy describing unconventional
superconductors in the absence of external magnetic
fields. This allows a more precise description of the
magnetic instabilities previously considered in these
systems. In particular, we derive new sufficient conditions
for the stability of both the homogeneous and inhomogeneous
equilibrium states.


\end{abstract}

In recent years, a number of authors have
 considered the possibility of magnetic instabilities in
unconventional superconductors. Within the framework of a
phenomenological Landau-Ginzburg description of 
these materials \cite{r1},
a multicomponent order
parameter can  give rise to a number of couplings
which make the system potentially unstable against
spatial modulations.
The stability of homogeneneous phases was first considered
by Zhitomirskii \cite{r2}. Soon after that, Palumbo, Muzikar 
and
Sauls \cite{r3} described a magnetic instability in the
absence of externally  applied magnetic fields, associated
with the creation of supercurrents due to spatial
distorsions of the order parameter. They found the
instability  by studying the change in the
free energy under small perturbations about the homogeneous
phase.

Using similar techniques,  Zhitomirskii \cite{r4}
later identified several possible inhomogeneous phases; he
also derived some necessary conditions for the
Landau-Ginzburg free energy  functional to be bounded from
below.
This is a crucial question because the existence of an
unstable mode, while automatically implying the
instability of the homogeneous phase,  does not guarantee
the existence of an inhomogeneous phase. It could very
well happen that the free energy is unbounded from below.

On the other hand, it is important to realize that sufficient 
conditions for
the {\it instability} of a given phase are not the same as 
sufficient conditions
for its {\it stability}. The standard analysis of small 
perturbations can
only provide information about the former. The latter can 
sometimes be derived
from a study of the minima of the potential provided the 
gradient terms are
positive definite. Both methods tend  to give  conservative 
bounds, which often
leave a region of phase space intractable. Our aim here is to 
bridge the gap
between the two and obtain information on the stability in 
these intermediate
regions.

In this paper we present a general method which can be
used to obtain new  {\it sufficient} stability conditions for
Landau-Ginzburg free energies describing unconventional
superconductors. The basic idea,  inspired by Bogomolnyi's
classical treatment of vortices \cite{r6}, 
 is to rewrite the free
energy in such a way that lower bounds to its value are
apparent. This is achieved by using the boundary
conditions to transform and recombine different terms
into perfect squares, so that positivity becomes obvious.
As we shall see, our results are complementary to previous 
studies based on the analysis of small perturbations to the 
homogeneous equilibrium states, which yield necessary
stability conditions. In some cases we are able to establish 
both
necessary {\it and} sufficient conditions, thus providing   
the exact location of the phase transition.

\bigskip
Following \cite{r3} and \cite{r4} we consider an order 
parameter
$V=(V_1,V_2)$ transforming according to the
two-dimensional $E_{1g}$ representation of the point group
$D_{6h}$, but our method easily generalizes to other
cases. We will consider z-independent configurations,
with $B=\partial_x A_y - \partial_yA_x$ and $\vec D =
\vec {\nabla} - i\vec{A}$. The free energy density can be
written  as
\begin{equation}\label{e1}
F = F_B +F_G+ F_P 
\end{equation}
where $F_B = {1 \over 2} B^2$ is the magnetic energy
contribution (in the absence of externally applied
field), $F_G$ depends on the gradients,
\begin{equation}\label{e2}
F_G = \kappa_1 {\bar D}_\alpha {\bar V}_\beta
D_{\alpha} V_\beta 
+ \kappa_2 {\bar D}_\alpha {\bar V}_\alpha
D_{\beta}
V_\beta +
\kappa_3 {\bar D}_\alpha {\bar V}_\beta
D_{\beta} V_\alpha
\end{equation} 
and $F_P$ is the potential
\begin{equation}\label{e3}
F_P = {1 \over 2}\beta ({\bar V} V - 1)^2 + {1 \over 2}
\sigma (i{\bar V} \times V)^2
\end{equation}
with ${\bar V} V ={\bar V_1} V_1 +{\bar V_2} V_2$, and
$i({\bar V}\times V) = i({\bar V_1} V_2 -{\bar  V_2}
V_1)$. Note that we have already performed several rescalings
 in order to bring the free energy into this particularly 
simple
form.
We must take $\beta>0$  and $\sigma > -\beta$ for
global stability (these are {\it necessary} conditions).

Let us first consider the homogeneous phases. 
Setting $DV = B=0$ we have to
distinguish two cases, depending on the value of $\sigma$:

 $\bullet$ For $\sigma >0$, $F_P$ is the sum
of two positive squares and takes its minimum value $F_P=0$ 
for $V$ real 
($i{\bar V}\times V = 0$) with ${\bar V} V=1$, i.e.,
\begin{equation}\label{e4}
V_0 = \pmatrix{\cos \theta \cr \rm{sin} \theta\cr}
\qquad,\qquad \sigma >0 
\end{equation}
This is sometimes referred to as the {\it time reversal 
symmetric} phase in the
literature.

$\bullet$ For $-\beta<\sigma<0$, \ (\ref{e3}) can be 
rewritten as
\begin{equation}\label{e5}
F_P = {1\over 2} (\beta+\sigma) \left({\bar V} V-{\beta\over
{\beta+\sigma}}\right)^2-{1\over 2}\sigma\left( ({\bar V} 
V)^2-
(i{\bar V} \times V)^2\right)+{1\over 2}{\sigma\beta\over
{\beta+\sigma}}
\end{equation}
and we have $F_P\geq {1 \over 2}{\sigma\beta \over
\beta+\sigma}$. $F_P$ attains its minimum value for
$|{\bar V}V|=|i{\bar V} \times V| = {\beta \over
\beta+\sigma}$, i.e., 
\begin{equation}\label{e6}
V_0 = {1 \over \sqrt 2}\left( {\beta \over
\sigma+\beta}\right)^{1\over 2}e^{i\theta} \pmatrix{1\cr
\pm i\cr} \qquad,\qquad \sigma <0
\end{equation}
also known as the {\it time reversal symmetry breaking} state.
Eqns. \ (\ref{e4}) and \ (\ref{e6}) with $\vec A=0$ represent 
two types of
homogeneous  phases. In order to study their stability, we
must take into account the behaviour of $F_G$. Consider
first configurations with $\vec A =0$. The contribution
to $F_G$ from a plane wave with $V_2=0$ and $\vec {k} =
(k_x,k_y)$ will be proportional to
\begin{equation}\label{e7}
( \kappa_{123} k_x^2 + \kappa_1 k_y^2 ) |V_1|^2
 \end{equation}
and this will be positive for $\kappa_{123} \equiv
\kappa_1 + \kappa_2 + \kappa_3 >0$ and $\kappa_1 >0$.
These {\it necessary} conditions for global stability
of $F$ have been considered in \cite{r3} and \cite{r4}, and 
in what follows 
they will be assumed to hold.

Consider now general configurations with $\vec A \neq
0$. $F_G$ can be written
\begin{mathletters}\label{e8}
\begin{equation}
F_G = \kappa_{123} (|D_1V_1|^2 + |D_2V_2|^2) +
\kappa_2( {\bar D}_1{\bar V}_1 D_2V_2 + D_1V_1  {\bar
D}_2{\bar V}_2)  \label {e8a}
\end{equation}
\begin{equation}
+ \kappa_1 (|D_1V_2|^2 + |D_2V_1|^2) + \kappa_3 (
 {\bar D}_1{\bar V}_2 D_2V_1 + D_1V_2  {\bar
D}_2{\bar V}_1) \label{e8b}
\end{equation}
\end{mathletters}
 and it is clear that \ (\ref{e8a})
and \ (\ref{e8b}) will be positive definite forms in 
$\{ D_1V_1,D_2V_2\}$
and $\{ D_1V_2,D_2V_1\}$ respectively if
\begin{equation}\label{e9}
|\kappa_2|< \kappa_{123} \qquad\qquad {\rm and}
\qquad\qquad |\kappa_3|< \kappa_1 
\end{equation}
If \ (\ref{e9}) is satisfied we have $F_G \geq 0$. Since the
minimum value of $F_P$ is obtained for homogeneous
configurations (with $\vec A = 0$), it is clear that \ 
(\ref{e9})
represents {\it sufficient} conditions for the stability
of the homogeneous phases. These conditions were given
without proof in \cite{r2}.

We will now show that the domain of existence of the homogeneous 
phase is in
fact larger than  conditions \ (\ref{e9}) suggest.
Consider the case $\sigma >0$. There are two key observations 
which allow us to
enlarge the region of stability given by \ (\ref{e9}). The 
first one is that, up
to integration by parts (which is allowed, for instance, by 
periodic boundary
conditions) \begin{eqnarray}
( {\bar D}_1{\bar V}_1 D_2V_2 + D_1V_1  {\bar
D}_2{\bar V}_2) - ( {\bar D}_1{\bar V}_2 D_2V_1 + D_1V_2  {\bar
D}_2{\bar V}_1)  =  i(\bar{V}\times V)\cdot B
\nonumber
\end{eqnarray} 
and we may write 
\begin{equation}\label{e10}
 F_G(\kappa_1,\kappa_2,\kappa_3) = F_G(\kappa_1,
\kappa_2-\delta \kappa , \kappa_3 + \delta\kappa) +
i\delta\kappa (\bar{V}\times V)\cdot B \ \ .
\end{equation}
The second observation is that the extra term in \ (\ref{e10}) 
can
be combined in the following way with $F_P$ and $F_B$:

\begin{eqnarray}\label{e11}
F_B+F_P+i\delta\kappa (\bar{V}\times V)\cdot B = {1 \over
2} [B+i\delta\kappa (\bar{V}\times V)]^2
\nonumber \\
+{1\over 2}\beta (\bar{V}V -1)^2 + {1 \over 2} (\sigma
-\delta\kappa^2)[i (\bar{V}\times V)]^2\ \ .
\end{eqnarray}

Since this is positive definite as long as $|\delta
\kappa|<\sqrt{\sigma}$, this freedom in the choice of
$\delta \kappa$ in \ (\ref{e10}) can be used to 
transform \ (\ref{e9}) into
\begin{equation}\label{e12}
|\kappa_2| < \kappa_{123} + \sqrt{\sigma} \qquad\qquad 
{\rm and}
\qquad\qquad 
|\kappa_3| < \kappa_{1} + \sqrt{\sigma}
 \end{equation}
Basically we have ``pushed'' the boundary of the
stability region given by \ (\ref{e9}) along vectors $\pm
(\sqrt{\sigma}, -\sqrt{\sigma})$. Within the
region given by \ (\ref{e12}), $F\geq 0$, and the minimum value
$F=0$ is attained by the homogeneous state \ (\ref{e4}). Thus 
the
homogeneous phase is stable if \ (\ref{e12}) is satisfied 

We have just shown the sufficient character of \ (\ref{e12}). 
Equations \ (\ref{e12}) are also {\it necessary} conditions. 
This
follows from the consideration of perturbations
$V=V_0+\delta V$ to the homogeneous state \ (\ref{e4}). Using
rotational symmetry to set $\theta = 0$, one finds that
$F$ is unstable against perturbations of the form:
\begin{mathletters}\label{e13}
\begin{equation}
\delta V \propto \pmatrix {0\cr i \cos ky \cr} \ ; \qquad
|\kappa_2| > \kappa_{123} + \sqrt { \sigma}
\label{e13a}
\end{equation}
\begin{equation}
\delta V \propto \pmatrix{0\cr i  \cos kx \cr} \ ; \qquad
|\kappa_3| > \kappa_{1} + \sqrt { \sigma}
\label{e13b}
\end{equation}
\end{mathletters}
These perturbations 
were considered by
Zhitomirskii \cite{r4}, who showed that (12) are
 necessary conditions for the stability of \ (\ref{e4}). 
The fact that 
these are identical with our new sufficient conditions means 
that 
eqs. \ (\ref{e12}) define exactly the domain of existence 
of the homogeneous
phase \ (\ref{e4}).
                                                                        
Now let us consider what happens outside this domain. The point 
is that finding
an instability of the homogeneous phase does not in itself 
prove the existence of
an inhomogeneous phase. To do this,  one  also has to  show 
that the free energy
remains bounded below, and this cannot be established by 
studying its changes
under small perturbations.

Notice, however, that our method provides such a proof:
 according
to \ (\ref{e5}) $F_P$ is bounded from below as long as $\sigma
>-\beta$, therefore we may choose 
$|\delta \kappa| < \sqrt{\beta+\sigma}$, and conclude
that inside the region defined by
\begin{equation}\label{e14}
|\kappa_2| < \kappa_{123} + \sqrt{\beta+ \sigma}, \
\qquad 
|\kappa_3| < \kappa_{1} + \sqrt{\beta+ \sigma}, \
\qquad \end{equation}
the total free energy density is bounded from below. This
is obviously true irrespective of the sign of $\sigma$, and 
in the case where
$\sigma>0$ it proves the existence of a stable inhomogeneous 
phase without
 resorting to numerical techniques.

Note that eqs. \ (\ref{e13a}) and \ (\ref{e13b}) can be 
described 
as ``longitudinal'' and
``transverse'' modulations respectively and  that for
$\kappa_3<-\kappa_1$ both perturbations are
simultaneously unstable. In that region we may thus
expect the occurrence of 2-D modulated solutions. This
seems to be confirmed by the numerical simulations in
\cite{r5}.

To summarize, we have suggested a new approach to the study 
of lower bounds on
the free energy of condensed matter systems inspired on 
Bogomolnyi's ideas. Our
basic observation is that integration by parts generally 
results in 
a mixing of the gradient, potential and magnetic
contributions to the free energy which can be exploited to 
obtain sufficient
(rather than necessary) conditions for stability.
This gives complementary results to the
standard analysis of small perturbations and we have shown 
here that in some
cases it is enough to establish domains of existence for  
homogeneous and
inhomogeneous phases. We illustrated the method in a model 
based on a
two-dimensional representation of $D_{6h}$, for which we 
obtained the following
new {\it sufficient} conditions:
 
For $\sigma > 0$ the homogeneous state \ (\ref{e4}) is stable 
if and only if
$|\kappa_2| < \kappa_{123} + \sqrt{\sigma}, \ |\kappa_3| < 
\kappa_{1} + \sqrt{
\sigma}$ .

For $\sigma > -\beta$ and $|\kappa_2| < \kappa_{123} 
+ \sqrt{\beta+ \sigma}, \ |\kappa_3| < \kappa_{1} +
\sqrt{\beta+ \sigma}$, irrespective of the sign of $\sigma$, 
 the free energy density is bounded
from below.

Two questions remain open: We don't know the exact domain of 
stability of the homogeneous phase \ (\ref{e6}) for $\sigma < 
0$. 
All we
can say is that, as long as eqs. \ (\ref{e9}) are satisfied, 
the homogeneous 
phase \ (\ref{e6}) is stable. But phase \ (\ref{e6}) is known 
to be stable beyond that domain,
and the transition to the inhomogeneous state is of first 
order \cite{r4}. 

The second question concerns the global stability of the free 
energy 
beyond the region defined by eqs. \ (\ref{e14}), which 
are only sufficient conditions.
Consideration of a single plane wave \cite{r4} imposes some 
constraints on the
coefficients of the free energy, but little is known about 
general 
configurations. There are, however, two situations where 
we know that eqs. \ (\ref{e14})
are also necessary conditions for the free energy to be 
bounded from below:

For $\beta=0$, conditions \ (\ref{e12}) 
and \ (\ref{e14}) are identical, and we can use a scaling 
argument. 
Beyond the domain
defined by eqs. \ (\ref{e14}), the free energy will
take a negative value $f$ for a modulation of the form \ 
(\ref{e13}). 
A similar configuration
with $V\to n V$, $\vec A\to n\vec A$, and $\vec k\to n\vec k$ 
will 
satisfy the
same boundary conditions for $n$ integer and will have a 
free energy $n^4 f$. 
Since $n$ can be arbitrarily large, 
it is obvious that the free energy is unbounded
from below.

Note that this argument does {\it not} apply for $\beta\neq 0$. 
But we have  studied  the system numerically and  found that, 
for $\kappa_{23}\equiv\kappa_2+\kappa_3=0$, the free energy 
becomes unbounded
from below as soon as \ (\ref{e14}) is violated. However, 
our results for $\kappa_{23}\neq 0$
 are inconclusive.

\section*{Acknowledgements}

We thank G. Volovik and M. Zhitomirskii for their help
locating references,  M.A. Valle, M.A. Go\~ni, I.L.
Egusquiza and P. Tiesinga for conversations; and K. Kuijken
for help with numerical simulations. A.A. wishes to thank
the Isaac Newton Institute and the University of Utrecht
for their hospitality. This work was partially supported 
by the Isaac Newton Institute (Cambridge, U.K.), by NSF
grant PHY-9309364, CICYT grant AEN-93-1435 and the
University of the Basque Country grant UPV-EHU
063.310-EB119/92.

\end{document}